\documentclass[aps,prl,reprint,superscriptaddress,nofootinbib,longbibliography,floatfix]{revtex4-2}

\usepackage[T1]{fontenc}
\usepackage[utf8]{inputenc}
\usepackage{amsmath,amssymb,mathtools}
\usepackage{microtype}
\usepackage{xcolor}
\usepackage[colorlinks=true,allcolors=blue!55!black]{hyperref}

\newcommand{\AdS}{\mathrm{AdS}}
\newcommand{\dd}{\mathrm{d}}
\newcommand{\cL}{\mathcal{L}}
\newcommand{\cM}{\mathcal{M}}
\newcommand{\cH}{\mathcal{H}}
\newcommand{\cS}{\mathcal{S}}
\newcommand{\cO}{\mathcal{O}}
\newcommand{\half}{\tfrac{1}{2}}

\hypersetup{
  pdftitle={Null String Holography in AdS from Local BRST Descent},
  pdfauthor={Sarthak Duary and Sourav Maji}
}

\begin{document}

% \title{Null string holography in AdS from local quantum descent}

 \title{Null string holography in AdS from local BRST descent}

\author{Sarthak Duary}
\thanks{Corresponding author.\\
sarthakduary@tsinghua.edu.cn}
\affiliation{Yau Mathematical Sciences Center, Tsinghua University, Beijing 100084, China}

\author{Sourav Maji}
\thanks{Corresponding author.\\
souravmaji@hri.res.in}
\affiliation{Harish-Chandra Research Institute, A CI of Homi Bhabha National Institute,
Chhatnag Road, Jhunsi, Prayagraj (Allahabad), Uttar Pradesh 211019, India}

\date{\today}

\begin{abstract}
We show that the radial sector of a Carroll-Weyl gauged null string in Poincar\'e $\mathrm{AdS}_D$ is locally trivial in BRST cohomology. In a quotient-adapted BFV basis, the radial coordinate, its conjugate
momentum, and the associated ghosts form a contractible quartet. An
explicit contracting homotopy extends this reduction to the full algebra of local spatial jets, leaving precisely the $(D-1)$-dimensional ILST
BRST complex. The same quartet carries zero net Virasoro central charge,
so that the anomaly condition of the parent theory, $D=27$, descends to the ILST condition $d=26$.  
\end{abstract}

\maketitle

\paragraph{Introduction.---}
In AdS holography, the radial coordinate has a special role, since evolution in the radial direction is closely tied to the scale dependence of the boundary theory \cite{Maldacena1998,GubserKlebanovPolyakov1998,Witten1998,AharonyEtAlReview}. Usually, the boundary description is extracted asymptotically by studying bulk fields near the conformal boundary. For a Carroll-Weyl gauged null string, we find a different realization. If the radial flow in the AdS Poincar\'e patch is gauged, then the radial coordinate should not correspond to an independent physical degree of freedom. Its dependence should therefore drop out of gauge-invariant quantities through BRST reduction, without requiring an explicit $z\to0$ limit. From this point of view, the relation between bulk and boundary can be understood directly as a consequence of gauge equivalence, rather than as an asymptotic limiting procedure.

This classical quotient is the first step. In the quantum theory, the radial coordinate comes with its conjugate momentum and the ghosts associated with the radial constraint, and local operators may also involve spatial derivatives of these fields. It is therefore not enough to show that the basic radial variables are gauge trivial. We should also show that the full radial sector is trivial in the local BRST cohomology. This reduction is consistent with the anomaly structure and the functional measure.

 Null strings provide a natural framework for addressing this question \cite{Schild1977,IsbergEtAl1994}. Their Carrollian world-sheet structure is closely connected to Galilean conformal symmetry \cite{BagchiGCA2013,BagchiWorldsheet2015}, ambitwistor constructions \cite{MasonSkinner2014,CasaliTourkine2016,CasaliHerfrayTourkine2017}, and boundary Carrollian dynamics \cite{BagchiEtAlOpenNull}. More recently, an additional Carroll-Weyl gauge symmetry was identified \cite{SheikhJabbariYavartanooOverlooked,SheikhJabbariYavartanooI,SheikhJabbariYavartanooII} and the corresponding classical quotient in Poincar\'e AdS was shown to have a projective-boundary interpretation \cite{NullHolography}. A related embedding-space construction yields null strings on lower-dimensional (A)dS backgrounds \cite{LindstromSundborg}. 
This naturally raises the quantum question of whether the same geometric reduction survives at the level of BRST cohomology.

Here we answer this question locally. Starting from the radial-adapted Carroll-Weyl constraint basis developed in the companion analysis~\cite{DuaryMajiADS}, we formulate the radial quotient in a BFV basis that separates the $\AdS_D$ theory into a transverse $(D-1)$-dimensional ILST complex and a radial quartet $(z,\gamma)\oplus(\beta,P_z)$. We construct an explicit contracting homotopy on the complete equal-time spatial-jet algebra and prove that every BRST-closed local operator of positive radial degree is exact. The same quartet cancels from the Virasoro anomaly, giving the structural relation $D=27\leftrightarrow d=26$, and a regulated canonical integral supplies an independent measure-level check. Our result is local. A full treatment of global radial sectors, zero modes, and the completed state space lies beyond the present analysis.

\paragraph{Radial Gauge Reduction.---}
The classical origin of the descent is transparent in Poincar\'e coordinates $X^A=(z,x^i)$,
\begin{equation}
 \dd s_G^2=\frac{L^2}{z^2}
 \left(\dd z^2+\eta_{ij}\dd x^i\dd x^j\right),
 \qquad z>0 .
 \label{eq:adsmetric}
\end{equation}
The radial vector $K=\partial_z$ is conformal Killing. In a local conformal
frame, defined by
\begin{equation}
 g_{AB}=z^2e^{2z}G_{AB},
 \qquad
 \mathcal L_{\partial_z}g_{AB}=2g_{AB},
\end{equation}
it becomes homothetic \cite{NullHolography}. Introducing the invariant fields
\begin{equation}
 U^a=e^zV^a,
 \qquad B_a=\partial_a z+W_a,
\end{equation}
the gauged configuration-space action becomes
\begin{equation}
 S_K=\frac{\kappa L^2}{2}\int_\Sigma\!\dd^2\sigma\,
 U^aU^b\left(B_aB_b+\eta_{ij}\partial_ax^i\partial_bx^j\right).
 \label{eq:invariantaction}
\end{equation}
Here $V^a$ is the world-sheet vector density and $W_a$ the Carroll-Weyl connection \cite{NullHolography}. On a regular patch with nonzero $U^a$, variation with respect to $W_a$ imposes $U^aB_a=0$. Substituting this condition removes the radial contribution and leaves the ILST action for the transverse coordinates. The classical quotient is therefore labelled by $x^i$; because each fixed-$x^i$ radial orbit approaches $z=0$, these labels may also be identified with coordinates on the Poincar\'e projective boundary \cite{NullHolography}. The quotient itself, however, is defined by the gauge orbit and does not require a $z\to0$ limiting procedure.

We now turn to the phase-space formulation, where the structure of the reduction becomes transparent. The canonical action is
\begin{equation}
 S_{\rm ph}=\frac{1}{2\pi}\int\!\dd\tau\dd\sigma\,
 \bigl[P_z\dot z+P_i\dot x^i-e\cM-u\cH-a\cS\bigr],
 \label{eq:phaseaction}
\end{equation}
with
\begin{equation}
 \cM=\half(P_z^2+P_iP^i),\qquad
 \cH=P_zz'+P_ix'^i,\qquad \cS=P_z .
 \label{eq:constraints}
\end{equation}
The last equality is essential. A gauged target-space flow generated by $K^A$ has canonical generator $K^AP_A$; hence $K=\partial_z$ generates translations of the radial coordinate, not dilations \cite{DuaryMajiADS}. This differs from the flat-space Carroll-Weyl constraint $\cS=X\!\cdot\!P$ \cite{DuaryMajiBRST,ChenHuAnomalies}.

The Hamiltonian constraint $\cM$ in Eq.~\eqref{eq:constraints} is the
radial-adapted representative introduced in Ref.~\cite{DuaryMajiADS}.
It is invariant along the Carroll-Weyl radial flow and represents the
same local classical constraint system as the natural homothetic
constraint. 
% A further quotient-adapted redefinition below separates
% the transverse constraints from the radial sector.

Writing $\cL=-\cH$, the residual modes obey
\begin{align}
 [L_m,L_n]&=(m-n)L_{m+n}, & [L_m,M_n]&=(m-n)M_{m+n},\nonumber\\
 [L_m,S_n]&=-nS_{m+n}, & [S_m,M_n]&=0,
 \label{eq:adsalgebra}
\end{align}
with $[M_m,M_n]=[S_m,S_n]=0$. Thus the residual algebra takes the form of a centerless $\mathrm{BMS}_3$ algebra extended by a weight-one abelian current $S$ \cite{DuaryMajiADS}.

For the quotient, we now introduce the equivalent adapted generators
\begin{align}
 \widehat{\cM}&=\cM-\half\cS^2=\half P_iP^i,\nonumber\\[-2pt]
 \widehat{\cL}&=\cL+z'\cS=-P_ix'^i,\nonumber\\[-2pt]
 \widehat{\cS}&=\cS=P_z .
 \label{eq:adaptedconstraints}
\end{align}
The transformation is invertible and preserves both the constraint surface and the radial gauge flow. In this form, the first two generators coincide with the ILST constraints in \((D-1)\)-dimensions, while the radial sector is carried entirely by \(\widehat{\cS}=P_z\).

The classical reduction is then straightforward. We choose the regular radial gauge
\begin{equation}
 \chi_z=z-z_\star=0,\qquad z_\star>0,
\end{equation}
for which
\begin{equation}
 \{\chi_z(\sigma),P_z(\sigma')\}
 =2\pi\delta(\sigma-\sigma').
\end{equation}
After gauge fixing, the pair $(\chi_z,P_z)$ becomes second class and can be eliminated. The transverse variables are unaffected, and their Dirac bracket remains canonical,
\begin{equation}
 \{x^i(\sigma),P_j(\sigma')\}_D
 =2\pi\delta^i{}_j\delta(\sigma-\sigma').
 \label{eq:diracbracket}
\end{equation}
The local configuration-space count agrees as well,
\begin{equation}
 N_{\rm conf}^{\rm parent}=D-3,
 \qquad
 N_{\rm conf}^{\rm ILST}=(D-1)-2,
 \label{eq:degreecount}
\end{equation}
so the two local configuration-space counts agree. Hence, locally, the classical quotient is the transverse ILST phase space. The field dependence of Eq.~\eqref{eq:adaptedconstraints} makes the corresponding BFV transformation nontrivial. It is therefore convenient to work directly in the regular BFV basis associated with the redefined constraints.

\paragraph{Local BRST Descent.---}
The redefined constraints make the separation between the transverse and radial
degrees of freedom explicit. The generators $\widehat{\cL}$ and
$\widehat{\cM}$ are precisely the ILST constraints in \((D-1)\)-dimensions,
whereas the radial direction is carried entirely by
$\widehat{\cS}=P_z$. In the BFV basis associated with these constraints, the BRST charge decomposes locally as
\begin{equation}
 Q_K=Q_{\rm ILST}^{(D-1)}+Q_{\rm rad},\qquad
 \{Q_{\rm ILST}^{(D-1)},Q_{\rm rad}\}=0,
 \label{eq:Qsplit}
\end{equation}
with
\begin{equation}
 Q_{\rm rad}
 =\frac{1}{2\pi}\int_0^{2\pi}\!\dd\sigma\,\gamma P_z .
 \label{eq:qrad}
\end{equation}
The corresponding radial BRST differential,
$s_{\rm rad}F:=\{Q_{\rm rad},F\}$, acts as
\begin{equation}
 s_{\rm rad}(z,\gamma,\beta,P_z)
 =(-\gamma,0,P_z,0),
 \label{eq:qrad-action}
\end{equation}
so $(z,\gamma)$ and $(\beta,P_z)$ form two BRST doublets. For local operators, it is not enough to consider only the undifferentiated quartet. One must also include all of its spatial derivatives, to arbitrary finite order \cite{BrandtJet2001}.

At fixed world-sheet time, we introduce the radial spatial-jet variables
\begin{equation}
\begin{aligned}
 z_{(n)}&=\partial_\sigma^n z, &
 \gamma_{(n)}&=\partial_\sigma^n\gamma,\\
 \beta_{(n)}&=\partial_\sigma^n\beta, &
 P_{z,(n)}&=\partial_\sigma^nP_z,
\end{aligned}
 \label{eq:radialjets}
\end{equation}
for $n\ge0$. We also define
$y_{(0)}=z-z_\star$ and $y_{(n)}=z_{(n)}$ for $n\ge1$.
We take $\mathcal A_{\rm loc}$ to be the algebra of local polynomials in the transverse and radial jets, with each element involving only finitely many jet variables. Since $s_{\rm rad}$ commutes with $\partial_\sigma$, its action on the
basic fields extends directly to the jet variables,
\begin{equation}
 s_{\rm rad}(y_{(n)},\gamma_{(n)},\beta_{(n)},P_{z,(n)})
 =(-\gamma_{(n)},0,P_{z,(n)},0).
 \label{eq:jetdoublets}
\end{equation}
To analyze the radial dependence of the local BRST complex, we introduce a radial counting operator $N_{\rm rad}$ and a contracting homotopy $h_{\rm rad}$ on $\mathcal A_{\rm loc}$ as follows
\begin{align}
 N_{\rm rad}=\sum_{n\ge0}\bigg(&
 y_{(n)}\frac{\partial}{\partial y_{(n)}}+
 \gamma_{(n)}\frac{\partial}{\partial\gamma_{(n)}}\nonumber\\[-2pt]
 &+\beta_{(n)}\frac{\partial}{\partial\beta_{(n)}}+
 P_{z,(n)}\frac{\partial}{\partial P_{z,(n)}}\bigg),\nonumber\\
 h_{\rm rad}=\sum_{n\ge0}\bigg(&
 -y_{(n)}\frac{\partial}{\partial\gamma_{(n)}}+
 \beta_{(n)}\frac{\partial}{\partial P_{z,(n)}}\bigg).
 \label{eq:homotopyoperators}
\end{align}
For Grassmann-odd variables, we use graded left derivatives
\cite{BarnichBrandtHenneaux2000}. Since a local polynomial involves only finitely many jet variables, both sums act finitely on every element of $\mathcal A_{\rm loc}$. From Eq.~\eqref{eq:jetdoublets}, it follows that
\begin{equation}
 \{s_{\rm rad},h_{\rm rad}\}=N_{\rm rad}.
 \label{eq:homotopyidentity}
\end{equation}
The transverse differential acts on disjoint variables, so for $s_K=s_{\rm ILST}^{(D-1)}+s_{\rm rad}$ one has
\begin{equation}
 [s_K,N_{\rm rad}]=0,\qquad
 \{s_K,h_{\rm rad}\}=N_{\rm rad}.
 \label{eq:fullhomotopy}
\end{equation}

Let $\cO_k$ have definite radial degree $k>0$ and satisfy $s_K\cO_k=0$. Then
\begin{equation}
 \cO_k=\frac1kN_{\rm rad}\cO_k
 =s_K\!\left(\frac1k h_{\rm rad}\cO_k\right),
 \label{eq:positiveexact}
\end{equation}
so every closed local operator of positive radial degree is exact. For a general local polynomial $\cO=\sum_{k=0}^{k_{\max}}\cO_k$, the condition $[s_K,N_{\rm rad}]=0$ allows us to treat each radial degree separately.
All components with $k>0$ are BRST exact, leaving only the degree-zero part
$\cO_0$ in the cohomology. This projection sets the radial jets to zero and
evaluates the representative on the regular slice $z=z_\star$, rather than
at the endpoint $z=0$.

This leaves only the transverse ILST sector in the local BRST cohomology,
\begin{equation}
 H^\bullet(\mathcal A_{\rm loc},s_K)
 \cong
 H^\bullet\!\left(\mathcal A_{\rm ILST},s_{\rm ILST}^{(D-1)}\right).
 \label{eq:cohomology}
\end{equation}
Thus the radial variables, their ghosts, and all of their spatial derivatives
do not contribute to the local BRST cohomology
\cite{KugoOjima1979,HenneauxTeitelboim1992,FischEtAl1989,BarnichBrandtHenneaux2000,BrandtJet2001,Quadri2002}.
The reduction therefore holds for the full local jet algebra, not only for the
undifferentiated BRST doublets.

The same reduction can be expressed directly at the level of the BRST complex. Let $\pi$ denote the projection onto radial degree zero, let $\iota$ be the natural inclusion of the transverse algebra, and define
\begin{equation}
 H\cO=\sum_{k=1}^{k_{\max}}\frac1k\,h_{\rm rad}\cO_k .
\end{equation}
These maps satisfy
\begin{equation}
 \pi\iota=1,\qquad
 s_KH+Hs_K=1-\iota\pi .
 \label{eq:deformationretract}
\end{equation}
It follows that every BRST class in the full localcomplex has a representative depending only on the transverse variables. Conversely, if a transverse operator is BRST exact in the full complex, it is already exact in the ILST subcomplex.

For a BRST-closed local observable
$V_K=\sum_{k=0}^{k_{\max}} V_{K,k}$, this relation can be written explicitly as
\begin{align}
 V_K
 &=
 V_{\partial}(x^i,P_i;\text{transverse ghosts})
 +s_K\Lambda,
 \label{eq:operatormap}\\
 \Lambda
 &=
 \sum_{k\ge1}\frac1k\,h_{\rm rad}V_{K,k}.
 \label{eq:lambdadef}
\end{align}
Here $V_{\partial}=\iota\pi V_K$ is the transverse representative, while all radial dependence is contained in the BRST-exact term $s_K\Lambda$. Thus local observables in the parent BRST complex are represented, up to BRST-exact terms, by observables of the $(D-1)$-dimensional ILST theory. 

\paragraph{Anomaly Descent: $27\to26$.---}
We now turn to the anomaly structure and ask whether the same reduction is reflected in the quantum anomaly structure. At fixed Carrollian time, we use the highest-weight first-order \cite{DuaryMajiBRST} OPE 
\begin{equation}
 P_A(w)X^B(v)\sim-\frac{\delta_A{}^B}{w-v},
 \label{eq:firstorderope}
\end{equation}
with matter currents
\begin{equation}
 T^X=-{:}P_A\partial X^A{:},\qquad
 M^X=\half{:}P_AP^A{:},\qquad
 S^X=P_z .
 \label{eq:mattercurrents}
\end{equation}

These are the radial-adapted currents of
Ref.~\cite{DuaryMajiADS}; the anomaly coefficients below refer to this
realization.

Although the abstract residual algebra admits independent $LL$, $LM$, $LS$, and $SS$ central cocycles \cite{DuaryMajiADS}, this realization generates only the Virasoro one,
\begin{equation}
 (c_{LL},c_{LM},c_{LS},c_{SS})_X=(2D,0,0,0).
 \label{eq:matterlevels}
\end{equation}
The first-order OPE has no $P_AP_B$ contraction. Consequently $M^X(w)M^X(v)$ and $S^X(w)S^X(v)$ have no central pole, while $T^X(w)M^X(v)$ and $T^X(w)S^X(v)$ contain only the standard weight-two and weight-one transformation poles.

The constraints $\mathcal L$ and $\mathcal M$ each give rise to a fermionic $(b,c)$ system of weights $(2,-1)$, while the abelian constraint $\mathcal S$ gives a fermionic pair of weights $(1,0)$. A fermionic first-order system of weights $(h,1-h)$ contributes \cite{DuaryMajiBRST}
\begin{equation}
 c=-2(6h^2-6h+1),
\end{equation}
so the total Virasoro ghost contribution is
\begin{equation}
 c_{LL}^{\rm gh}=-26-26-2=-54.
\end{equation}
In the adapted basis associated with the hatted constraints, the transverse ILST ghost system carries no independent $LM$ cocycle, and the radial scalar ghost enters through $T_{\rm rad}^{\rm gh}=-{:}\beta\partial\gamma{:}$. Thus
\begin{equation}
 (c_{LL},c_{LM},c_{LS},c_{SS})_{\rm gh}=(-54,0,0,0),
\end{equation}
and hence
\begin{equation}
 (c_{LL},c_{LM},c_{LS},c_{SS})_{\rm tot}=(2D-54,0,0,0).
 \label{eq:totallevels}
\end{equation}

The adapted decomposition makes the cancellation transparent,
\begin{equation}
 2D-54=
 \underbrace{2(D-1)-52}_{\text{transverse ILST}}
 +\underbrace{2-2}_{\text{radial quartet}} .
 \label{eq:anomalysplit}
\end{equation}
The first-order pair $(z,P_z)$ contributes $+2$, while the scalar ghost pair $(\beta,\gamma)$ contributes $-2$. The radial sector that drops out of the local BRST cohomology also carries no net Virasoro central charge. 

This cancellation is visible directly in the radial currents,
\begin{align}
 T_{\rm rad}&=-{:}P_z\partial z{:}-{:}\beta\partial\gamma{:}
 =s_{\rm rad}\!\left(-{:}\beta\partial z{:}\right),\nonumber\\
 M_{\rm rad}&=\half{:}P_z^2{:}
 =s_{\rm rad}\!\left(\half{:}\beta P_z{:}\right),
 \label{eq:exactcurrents}
\end{align}

which are themselves BRST exact in the adapted complex. The anomaly condition of the parent theory therefore reduces precisely to that of the $(D-1)$-dimensional ILST theory,
\begin{equation}
 D=27\qquad\Longleftrightarrow\qquad d=D-1=26.
 \label{eq:criticalmatch}
\end{equation}
Equation~\eqref{eq:criticalmatch} is specific to the highest-weight realization used here. It shows that the anomaly-cancellation conditions of the parent and reduced theories agree in this representation. Other choices of vacuum, improvement terms, or regularization require a separate analysis, including the allowed central extensions together with zero-mode and normal-ordering contributions \cite{DuaryMajiBRST,BagchiThreeVacua2020}.

\paragraph{Path Integral Descent.---}
The radial reduction can also be seen directly in the canonical path integral of the radial-adapted formulation. Using
\begin{equation}
\cM=\widehat{\cM}+\half P_z^2,
\qquad
\cH=-\widehat{\cL}+z'P_z,
\end{equation}
we shift the multiplier of the radial constraint according to
\begin{equation}
a'=a+\half eP_z+uz' .
\label{eq}
\end{equation}
To define the path integral carefully, we first regulate it by a finite-mode truncation or an equivalent lattice discretization. In this regulated setting, the shift has unit Jacobian, and the radial part of the action reduces to
\begin{equation}
 S_{\rm rad}
 =
 \frac{1}{2\pi}\int\!\dd\tau\,\dd\sigma\,
 P_z(\dot z-a').
 \label{eq:radialaction}
\end{equation}

Choosing the regular gauge $z=z_\star$ gives a field-independent Faddeev-Popov determinant. The regulated radial integral then takes the form
\begin{equation}
\begin{aligned}
 I_{\rm rad}
 &=\int\mathcal D a'\,\mathcal D P_z\,\mathcal D z\,
 \delta[z-z_\star]\,
 e^{\frac{i}{2\pi}\int P_z(\dot z-a')}\\
 &\propto\int\mathcal D P_z\,\mathcal D z\,
 \delta[P_z]\delta[z-z_\star]
 ={\cal N}_\Sigma .
\end{aligned}
 \label{eq:radialintegral}
\end{equation}
Here ${\cal N}_\Sigma$ is independent of the transverse fields. Integrating out the radial sector therefore leaves
\begin{equation}
 S_{\rm reduced}
 =
 \frac{1}{2\pi}\int\!\dd\tau\,\dd\sigma\,
 \left[P_i\dot x^i-\half eP_iP^i-uP_ix'^i\right],
 \label{eq:reducedaction}
\end{equation}
with the transverse canonical measure. The regulated path integral thus reduces as
\begin{equation}
 \mathcal D\mu_{\rm CW}^{(D)}
 \longrightarrow
 {\cal N}_\Sigma\,
 \mathcal D\mu_{\rm ILST}^{(D-1)} .
 \label{eq:measuredescent}
\end{equation}
For local BRST observables that do not depend on the radial quartet, the regulated path integral becomes
\begin{equation}
 Z_{K,\Sigma}\!\left[\prod_a\cO_a\right]
 ={\cal N}_\Sigma\,
 Z_{{\rm ILST},\Sigma}\!\left[\prod_a\cO_a\right].
 \label{eq:partitionrelation}
\end{equation}
Since ${\cal N}_\Sigma$ is field independent, it cancels from normalized expectation values,
\begin{equation}
 \left\langle\prod_a\cO_a\right\rangle_{K,\Sigma}
 =
 \left\langle\prod_a\cO_a\right\rangle_{{\rm ILST},\Sigma}.
 \label{eq:normalizedcorrelators}
\end{equation}
This is the path-integral counterpart of the operator relation in
Eq.~\eqref{eq:operatormap}. Observables of the reduced ILST theory are obtained simply by restricting to operators that are independent of the radial quartet. Conversely, every local BRST class in the parent complex can be represented in this form up to a BRST-exact term. The canonical path integral thus provides an independent check of the same local reduction, without using the OPE analysis \cite{DuaryMajiPI,DuaryMajiBRST}. Extending the argument to the continuum requires additional care with the global gauge slice, zero modes, and possible boundary contributions in field space.

\paragraph{Conclusions and Outlook.---}
We have shown that the Carroll-Weyl gauged null string in Poincar\'e $\AdS_D$ admits a local BRST reduction to the $(D-1)$-dimensional ILST theory. In a BFV basis, adapted to the radial quotient, the radial coordinate, its conjugate momentum, the associated ghosts, and all of their spatial derivatives form a contractible sector. As a result, the local BRST cohomology is entirely represented by the transverse variables and coincides with that of the $(D-1)$-dimensional ILST theory.

The same reduction is reflected in the anomaly structure. In the highest-weight realization used here, the radial matter and ghost contributions cancel in the Virasoro central charge, relating the parent condition at $D=27$ to the ILST condition at $d=26$. The same reduction is also reflected in the canonical path integral.
After fixing the radial gauge, the radial sector contributes only a
field-independent factor, and the remaining functional integral reduces
to that of the transverse ILST theory.

This reduction is different from simply taking the boundary limit $z\to0$. The BRST quotient is defined on a regular slice $z=z_\star>0$, whereas the conformal boundary enters only in the geometric interpretation of the reduced theory. Since radial orbits at fixed $x^i$ approach $z=0$, the surviving transverse variables can be naturally identified with the corresponding projective boundary data.

Several questions remain open. A complete quantum analysis will require a careful treatment of radial zero modes, global boundary conditions, and the full physical state space. An important next step is to understand the anomaly matching beyond the
highest-weight realization and to extend the local correspondence to
integrated vertex operators, correlation functions, and higher-genus
amplitudes. We return to these questions in future works. It would also be interesting to explore whether analogous BRST descents
occur in other backgrounds and in supersymmetric extensions.

\paragraph{Acknowledgments.---}
We thank M.~M.~Sheikh-Jabbari for his valuable comments on our draft. SD is supported by the Shuimu Tsinghua Scholar Program of Tsinghua University and the Beijing Natural Science Foundation of China under Grant No.~IS25035.

\bibliography{references}

\end{document}